\documentclass{webofc}

\usepackage[varg]{txfonts}   
\usepackage{hyperref}
\usepackage{url}
\hypersetup{colorlinks=true,citecolor=blue,urlcolor=blue,linkcolor=blue}

\usepackage{dcolumn} \usepackage{bm} \usepackage{amsfonts} \usepackage{amsmath} \usepackage{amssymb} 

\newcommand{\h}[1]{\widehat{#1}}

\newcommand{\de}{\partial}

\renewcommand{\tilde}[1]{{\widetilde{#1}}}

\newcommand{\group}[1]{\relax\ifmmode\mathsf{#1}\else\textsf{#1}\fi}

\newcommand{\WFA}{\mathcal{A}}

\newcommand{\mf}[1]{\mathfrak{#1}}

\begin{document}
\title{Dissipative contributions to spin polarization}
	
\author{Matteo Buzzegoli\fnsep\thanks{\email{matteo.buzzegoli@e-uvt.ro}}}
	
\institute{Department of Physics, West University of Timişoara, Bd. Vasile Pârvan 4, Timişoara 300223, Romania}
	
\abstract{%
%
%
Utilizing the Zubarev non-equilibrium density operator method and a first-order gradient expansion, we provide a comprehensive classification of all possible spin polarization effects. A key finding is that, with the unique exception of those induced by the gradient of the spin potential, all first-order dissipative corrections are chiral, implying their dependence on parity-breaking interactions or chiral imbalance. The work also introduces new non-dissipative phenomena akin to the Spin Hall Effect, including the Chiral Spin Hall Effect, which offers a novel observable to probe QCD topological configurations.
}
\maketitle
	
\section{Introduction}
Spin polarization measurements in relativistic heavy-ion collisions offer crucial insights into the properties of the quark-gluon plasma (QGP)~\cite{Becattini:2024uha}. Initial theoretical predictions~\cite{Becattini:2024uha,Becattini:2015ska}, based on the assumption of local thermodynamic equilibrium for spin degrees of freedom, indicated that thermal vorticity (fluid rotation and acceleration fields) should induce spin polarization. Measurements of global spin polarization of $\Lambda$ and $\bar{\Lambda}$ hyperons, and later $\Xi$ and $\Omega$ hyperons, confirmed these predictions~\cite{STAR:2017ckg,STAR:2021beb,STAR:2020xbm,ALICE:2019onw}.
	
However, subsequent measurements of local spin polarization (momentum-dependent polarization)~\cite{STAR:2019erd,ALICE:2021pzu} presented significant discrepancies with theoretical predictions, particularly an opposite sign for the longitudinal component, i.e. along the beam direction. This led to theoretical investigations incorporating other hydrodynamic fields, such as the thermal shear tensor~\cite{Liu:2021uhn,Becattini:2021suc} and gradients of chemical potential~\cite{Liu:2020dxg}, which improved agreement with experimental data. These observations also raised the question of whether spin degrees of freedom in the QGP are fully equilibrated~\cite{Hongo:2022izs,Wagner:2024fhf}. Addressing this necessitates the development of spin hydrodynamics~\cite{Florkowski:2017ruc,Weickgenannt:2019dks}, a framework that includes the evolution of a spin tensor and potentially a spin potential. This work focuses on comprehensively classifying all possible contributions to spin polarization in a dissipative fluid, up to the first order in a gradient expansion.
	
\section{Theoretical Framework}
Our analysis is grounded in relativistic quantum statistical mechanics, employing the Zubarev method~\cite{Zubarev:1979,Becattini:2014yxa} to describe an out-of-equilibrium system that reaches local thermal equilibrium on a spacelike hypersurface. The system's state is characterized by fundamental operators: the energy-momentum tensor $\widehat{T}^{\mu\nu}$, spin tensor $\widehat{S}^{\lambda,\mu\nu}$, conserved current $\widehat{j}^\mu$, and axial current $\widehat{j}_A^\mu$. By maximizing entropy under these fundamental constraints, a statistical operator is derived:
\begin{multline}
	\label{eq:StatOper}
	\h{\rho} = \frac{1}{Z} \exp\Bigg\{ 
	-\int_{\Sigma} {\rm d}\Sigma_\mu(y)\left(\h{T}^{\mu\nu}(y)\beta_\nu(y)
	-\zeta(y)\,\h{j}^\mu(y) -\zeta_A(y)\,\h{j}_A^\mu(y)
	-\frac{1}{2}\mf{S}_{\lambda\nu}(y)\h{S}^{\mu\lambda\nu}(y)\right) \\ +
	\int_\Omega {\rm d}\Omega\left[\h{T}_S^{\mu\nu}\xi_{\mu\nu} + \h{T}_A^{\mu\nu}\left(\mf{S}_{\mu\nu}-\varpi_{\mu\nu}\right)
	-\h{j}^\mu \nabla_\mu \zeta - \nabla_\mu\left(\zeta_A \h{j}_A^\mu \right) -\frac{1}{2}\h{S}^{\mu\lambda\nu} \nabla_\mu \mf{S}_{\lambda\nu} \right]\Bigg\},
\end{multline}
where $\Sigma$ is a 3D hypersurface where the particle hadronize, $\Omega$ is the four-volume encompossing $\Sigma$ and the initial time hypersurface, and $\h{T}_S^{\mu\nu}$ and $\h{T}_A^{\mu\nu}$  denote respectively the symmetric and anti-simmetric part of the energy momentum tensor.
We then apply a gradient expansion to this operator, focusing on terms up to the first order in derivatives of hydrodynamic fields, including the four-temperature vector $\beta^\mu = u^\mu /T$, the chemical potentials $\zeta = \mu/T$ and $\zeta_A = \mu_A/T$, and the spin potential $\mathfrak{S}^{\mu\nu}$. These hydrodynamic fields and their gradients are decomposed into their irreducible components under SO(3) symmetry respect to the fluid velocity $u$ to systematically identify all possible contributions.

The spin polarization $S^\mu$ of fermions of definite momentum $k$ is directly derived from the axial ($\mathcal{A}^\mu$) and scalar  ($\mathcal{F}$) parts of the Wigner function $W$~\cite{Becattini:2024uha}:
\begin{equation}
	\label{eq:SMass}
	S^\mu(k)= \frac{1}{2}\frac{\int_\Sigma {\rm d}\Sigma\cdot k\; \WFA_+^\mu(x,k)}
	{\int_\Sigma {\rm d}\Sigma\cdot k\; \mathcal{F}_+(x,k)}= \frac{1}{2}\frac{\int_\Sigma {\rm d}\Sigma\cdot k\; {\rm Tr}_4\left[\gamma^\mu \gamma^5 W_+(x,k)\right]}
	{\int_\Sigma {\rm d}\Sigma\cdot k\; {\rm Tr}_4\left[ W_+(x,k)\right]},
\end{equation}
where the subscript $+$ denote the particle part of the Wigner function.
Linear response theory is employed to systematically account for corrections to the mean value of an observable, in this context, the Wigner operator $\widehat{W}(x,k)$.  A distinguishing feature of our methodology is the derivation of momentum-dependent Kubo formulas~\cite{Liu:2021uhn,Liu:2020dxg,Buzzegoli:2025zud}, which precisely define the thermal and transport coefficients associated with these induced effects. These coefficients are rigorously classified based on their transformation properties under discrete symmetries: parity (P), time-reversal (T), and charge conjugation (C), allowing for a clear distinction between non-dissipative and dissipative, and non-chiral and chiral contributions.

\section{Results and Discussion}
We have comprehensively classified all first-order gradient contributions to the axial part of the Wigner function. These contributions are categorized into non-dissipative and dissipative components, further subdivided into non-chiral and chiral categories. We introduce the notation:
%
$
\Delta^{\mu\nu} = \eta^{\mu\nu} - u^\mu u^\nu,\,
k_\perp^{\rho} =\Delta^\rho_\lambda k^\lambda,\,
Q^{\mu\nu} = \frac{\Delta^{\mu\nu}}{3} - \frac{k_\perp^\mu k_\perp^\nu}{k_\perp^2}.
$
%

\emph{Non chiral non-dissipative contributions} include previously identified effects such as spin polarization induced by the thermal shear, a chiral imbalance and the spin Hall effect originating from gradients of the vector chemical potential. The thermal vorticity $\varpi_{\mu\nu} = \alpha_\mu u_\nu - \alpha_\nu u_\mu + \epsilon_{\mu\nu\rho\sigma}w^\rho u^\sigma$ induces a polarization that persist even at global thermal equilibrium, and our findings indicate that interactions can break the degeneracies observed in free fields, where $a_{w u} = a_{w \Delta}$, $a_{\alpha\epsilon} = a_{w \Delta}$ and  $ a_{w k}=0$ in
\begin{equation}
	\label{eq:DWFALTEvort}
	\Delta_{{\rm LTE},\varpi} \WFA_+^\mu = -a_{\varpi} \frac{2\tilde\varpi^{\mu\nu}k_\nu}{(2\pi)^3}
	- (a_{w u} - a_{w \Delta}) \frac{(k\cdot w)u^\mu}{(k\cdot u)} 
	+ (a_{\alpha\epsilon} - a_{w \Delta})  \frac{\epsilon^{\mu\nu\rho\sigma} k^\perp_\nu u_\sigma \alpha_\rho}{(k\cdot u)} + a_{w k} Q^{\mu\rho} w_\rho ,
\end{equation}
thereby leading to additional contributions beyond what is seen in non-interacting systems. 
\emph{Chiral non-dissipative contributions} arise from the spin potential and gradients of the axial chemical potential and are detailed in~\cite{Buzzegoli:2025zud}. 

For the \emph{dissipative contributions}, our work identifies these effects for the first time in such a comprehensive manner. A pivotal finding is that \emph{all dissipative effects contributing to spin polarization, apart from those specifically arising from the gradients of the spin potential, fundamentally require the presence of a chiral imbalance or parity-violating interactions to be non-zero}. This implies that many dissipative effects might be negligible in heavy-ion collisions. However, if the spin degrees of freedom are not fully equilibrated, the gradients of the spin potential could play a significant role in observed phenomena.
It is important to note that many of these effects, with the exceptions of those derived from thermal vorticity and gradients of vector and axial chemical potential, exhibit dependence on the chosen pseudo-gauge.
	
This analysis also reveals the presence of new non-dissipative contributions:
\begin{equation}
	\Delta_{\rm SHE} \mathcal{A}_+^\mu(x,k)=\epsilon^{\mu\nu\rho\sigma}\frac{k^\perp_\nu u_\sigma}{(k\cdot u)}\left[ a^c_{r\epsilon}(k)\de_\rho\zeta + \mathfrak{a}_{r_A\epsilon}(k)\de_\rho\zeta_A \right],
\end{equation}
identifying and detailing a chiral version of the Spin Hall Effect (CSHE) $a^c_{r\epsilon}$, that is  a non-dissipative, non-chiral effect where spin polarization is induced by gradients of the chemical potential $\de_\rho\zeta$.
Instead, the Chiral Spin Hall Effect (CSHE) $\mathfrak{a}_{r_A\epsilon}$ is a new non-dissipative, chiral effect where spin polarization is induced by gradients of axial chemical potential $\de_\rho\zeta_A$.
These ``Hall-like'' effects are local phenomena; they appear only in local spin polarization and do not generate a net axial or vector current, meaning they are not detectable in global spin polarization.
Similarly, in the vector part of the Wigner function:
\begin{equation}
	\Delta_{\rm SHE} \mathcal{V}_+^\mu(x,k)=\epsilon^{\mu\nu\rho\sigma}\frac{k^\perp_\nu u_\sigma}{(k\cdot u)}\left[ \mathfrak{v}_{r\epsilon}(k)\de_\rho\zeta + v^c_{r_A\epsilon}(k)\de_\rho\zeta_A \right],
\end{equation}
the gradients of the (vector) chemical potential induce a non-dissipative  chiral effect,  the Chiral Electrical Effect (CEE) $ \mathfrak{v}_{r\epsilon}$, and the axial chemical potential induce a non-dissipative non-chiral  effect in the vector part of the Wigner function: the Axial Hall Effect (AHE) $v^c_{r_A\epsilon}$.
	
\section{Phenomenological Implications and Outlook}
The findings (see \cite{Buzzegoli:2025zud} for the complete results) have direct implications for understanding spin phenomena in various collision systems.
In heavy-ion collision at lower energies, the longer spin relaxation time suggests that spin degrees of freedom may not fully equilibrate. The inclusion of the spin potential and its gradients, as explored in this work, provides a framework for more accurate spin polarization predictions beyond local thermal equilibrium descriptions. Moreover, dissipative effects are expected to be more prominent in p-Pb collisions due to the inherently more out-of-equilibrium nature of the medium.
Finally, the Chiral Spin Hall Effect offers a novel observable to probe the topological configurations of Quantum Chromodynamics (QCD). This effect, induced by gradients of temperature and axial chemical potential in a direction orthogonal to both the particle's momentum and the gradients, provides a direct link between macroscopic spin observables and underlying topological charge fluctuations, especially in the presence of chiral imbalance from sphaleron transitions.

Future work will focus on estimating the phenomenological impact of these dissipative contributions by computing the transport coefficients. The presence of interactions could also lead to additional contributions even at local thermal equilibrium.


\begin{thebibliography}{18}
	
	\bibitem{Becattini:2024uha}
	F.~Becattini, M.~Buzzegoli, T.~Niida, S.~Pu, A.H. Tang, Q.~Wang, {Spin
		polarization in relativistic heavy-ion collisions}, Int. J. Mod. Phys. E
	\textbf{33}, 2430006 (2024), \texttt{2402.04540}.
	\doiwoc{10.1142/S0218301324300066}
	
	\bibitem{Becattini:2015ska}
	F.~Becattini, G.~Inghirami, V.~Rolando, A.~Beraudo, L.~Del~Zanna, A.~De~Pace,
	M.~Nardi, G.~Pagliara, V.~Chandra, {A study of vorticity formation in high
		energy nuclear collisions}, Eur. Phys. J. C \textbf{75}, 406 (2015),
	[Erratum: Eur.Phys.J.C 78, 354 (2018)], \texttt{1501.04468}.
	\doiwoc{10.1140/epjc/s10052-015-3624-1}
	
	\bibitem{STAR:2017ckg}
	L.~Adamczyk et~al. (STAR), {Global $\Lambda$ hyperon polarization in nuclear
		collisions: evidence for the most vortical fluid}, Nature \textbf{548}, 62
	(2017), \texttt{1701.06657}. \doiwoc{10.1038/nature23004}
	
	\bibitem{STAR:2021beb}
	M.S. Abdallah et~al. (STAR), {Global $\Lambda$-hyperon polarization in Au+Au
		collisions at $\sqrt {s_{NN}}$=3~GeV}, Phys. Rev. C \textbf{104}, L061901
	(2021), \texttt{2108.00044}. \doiwoc{10.1103/PhysRevC.104.L061901}
	
	\bibitem{STAR:2020xbm}
	J.~Adam et~al. (STAR), {Global Polarization of $\Xi$ and $\Omega$ Hyperons in
		Au+Au Collisions at $\sqrt {s_{NN}}$ = 200 GeV}, Phys. Rev. Lett.
	\textbf{126}, 162301 (2021), [Erratum: Phys.Rev.Lett. 131, 089901 (2023)],
	\texttt{2012.13601}. \doiwoc{10.1103/PhysRevLett.126.162301}
	
	\bibitem{ALICE:2019onw}
	S.~Acharya et~al. (ALICE), {Global polarization of $\Lambda \bar \Lambda$
		hyperons in Pb-Pb collisions at $\sqrt {s_{NN}}$ = 2.76 and 5.02 TeV}, Phys.
	Rev. C \textbf{101}, 044611 (2020), [Erratum: Phys.Rev.C 105, 029902 (2022)],
	\texttt{1909.01281}. \doiwoc{10.1103/PhysRevC.101.044611}
	
	\bibitem{STAR:2019erd}
	J.~Adam et~al. (STAR), {Polarization of $\Lambda$ ($\bar{\Lambda}$) hyperons
		along the beam direction in Au+Au collisions at $\sqrt{s_{_{NN}}}$ = 200
		GeV}, Phys. Rev. Lett. \textbf{123}, 132301 (2019), \texttt{1905.11917}.
	\doiwoc{10.1103/PhysRevLett.123.132301}
	
	\bibitem{ALICE:2021pzu}
	S.~Acharya et~al. (ALICE), {Polarization of $\Lambda$ and $\bar \Lambda$
		Hyperons along the Beam Direction in Pb-Pb Collisions at $\sqrt
		{s_{NN}}$=5.02\,\,TeV}, Phys. Rev. Lett. \textbf{128}, 172005 (2022),
	\texttt{2107.11183}. \doiwoc{10.1103/PhysRevLett.128.172005}
	
	\bibitem{Liu:2021uhn}
	S.Y.F. Liu, Y.~Yin, {Spin polarization induced by the hydrodynamic gradients},
	JHEP \textbf{07}, 188 (2021), \texttt{2103.09200}.
	\doiwoc{10.1007/JHEP07(2021)188}
	
	\bibitem{Becattini:2021suc}
	F.~Becattini, M.~Buzzegoli, A.~Palermo, {Spin-thermal shear coupling in a
		relativistic fluid}, Phys. Lett. B \textbf{820}, 136519 (2021),
	\texttt{2103.10917}. \doiwoc{10.1016/j.physletb.2021.136519}
	
	\bibitem{Liu:2020dxg}
	S.Y.F. Liu, Y.~Yin, {Spin Hall effect in heavy-ion collisions}, Phys. Rev. D
	\textbf{104}, 054043 (2021), \texttt{2006.12421}.
	\doiwoc{10.1103/PhysRevD.104.054043}
	
	\bibitem{Hongo:2022izs}
	M.~Hongo, X.G. Huang, M.~Kaminski, M.~Stephanov, H.U. Yee, {Spin relaxation
		rate for heavy quarks in weakly coupled QCD plasma}, JHEP \textbf{08}, 263
	(2022), \texttt{2201.12390}. \doiwoc{10.1007/JHEP08(2022)263}
	
	\bibitem{Wagner:2024fhf}
	D.~Wagner, M.~Shokri, D.H. Rischke, {Damping of spin waves}, Phys. Rev. Res.
	\textbf{6}, 043103 (2024), \texttt{2405.00533}.
	\doiwoc{10.1103/PhysRevResearch.6.043103}
	
	\bibitem{Florkowski:2017ruc}
	W.~Florkowski, B.~Friman, A.~Jaiswal, E.~Speranza, {Relativistic fluid dynamics
		with spin}, Phys. Rev. C \textbf{97}, 041901 (2018), \texttt{1705.00587}.
	\doiwoc{10.1103/PhysRevC.97.041901}
	
	\bibitem{Weickgenannt:2019dks}
	N.~Weickgenannt, X.L. Sheng, E.~Speranza, Q.~Wang, D.H. Rischke, {Kinetic
		theory for massive spin-1/2 particles from the Wigner-function formalism},
	Phys. Rev. D \textbf{100}, 056018 (2019), \texttt{1902.06513}.
	\doiwoc{10.1103/PhysRevD.100.056018}
	
	\bibitem{Zubarev:1979}
	D.N. Zubarev, A.V. Prozorkevich, S.A. Smolyanskii, Derivation of nonlinear
	generalized equations of quantum relativistic hydrodynamics, Theoretical and
	Mathematical Physics \textbf{40}, 821 (1979). \doiwoc{10.1007/BF01032069}
	
	\bibitem{Becattini:2014yxa}
	F.~Becattini, L.~Bucciantini, E.~Grossi, L.~Tinti, {Local thermodynamical
		equilibrium and the beta frame for a quantum relativistic fluid}, Eur. Phys.
	J. C \textbf{75}, 191 (2015), \texttt{1403.6265}.
	\doiwoc{10.1140/epjc/s10052-015-3384-y}
	
	\bibitem{Buzzegoli:2025zud}
	M.~Buzzegoli, {Kubo formulas for spin polarization in dissipative relativistic
		spin hydrodynamics: a first-order gradient expansion approach}, JHEP
	\textbf{7}, 255 (2025), \texttt{2502.15520}. \doiwoc{10.1007/jhep07(2025)255}
	
\end{thebibliography}
\end{document}